\documentclass[pra,superscriptaddress,a4paper,twocolumn]{revtex4-1}
\usepackage[english]{babel}
\usepackage[latin1]{inputenc}
\usepackage{hyperref}
\usepackage{amsmath, amssymb, amsfonts, mathrsfs}
\usepackage{graphicx}
\usepackage{algorithm}
\usepackage{algorithmic}
\usepackage{color}

\def\pmx{\begin{pmatrix}}
\def\emx{\end{pmatrix}}

\newcommand{\K}{\mathcal{K}}
\newcommand{\Hilb}{\mathcal{H}}
\newcommand{\U}{\mathcal{U}}
\newcommand{\pr}{^{\prime}}

\newcommand{\ket}[1]{|#1\rangle}

\newcommand{\card}[1]{\vert #1 \vert}

\begin{document} 

\title{Layered Quantum Key Distribution}

\author{Matej Pivoluska}
\affiliation{Institute for Quantum Optics and Quantum Information,
Austrian Academy of Sciences, Boltzmanngasse 3, Vienna A-1090, Austria}
\affiliation{Institute of Computer Science, Masaryk University, Botanick\'{a} 68a, 60200 Brno, Czech Republic}
\affiliation{Institute of Physics, Slovak Academy of Sciences,
D\'{u}bravsk\'{a} cesta 9,
845 11 Bratislava, Slovakia}

\author{Marcus Huber}
\affiliation{Institute for Quantum Optics and Quantum Information,
Austrian Academy of Sciences, Boltzmanngasse 3, Vienna A-1090, Austria}

\author{Mehul Malik}
\affiliation{Institute for Quantum Optics and Quantum Information,
Austrian Academy of Sciences, Boltzmanngasse 3, Vienna A-1090, Austria}

\begin{abstract}
We introduce a family of QKD protocols for distributing shared random keys within a network of $n$ users. The advantage of these protocols is that any possible key structure needed within the network, including broadcast keys shared among subsets of users, can be implemented by using a particular multi-partite high-dimensionally entangled quantum state. This approach is more efficient in the number of quantum channel uses than conventional quantum key distribution using bipartite links. Additionally, multi-partite high-dimensional quantum states are becoming readily available in quantum photonic labs, making the proposed protocols implementable using current technology.
\end{abstract}
\maketitle
\section{Introduction}
The possibility of increasing the amount of shared random variables across spatially separated parties in an intrinsically secure fashion is one of the flagship applications of quantum entanglement \cite{Ekert:1991kl}. Referred to as quantum key distribution (QKD), such schemes have matured to the point of commercial application today \cite{Clavis3}. 
Bipartite entanglement of a sufficient quality for violating Bell's inequalities is enough to ensure complete device-independent security in two-party communication scenarios 
\cite{Acin:2007db,Pironio:2009fd,Masanes:2011bv}. However, due to strict technical requirements such as extremely high detection and coupling efficiencies, such schemes are difficult to realize in practice \cite{Seshadreesan:2016bu}.

While conventional entanglement-based QKD protocols employ two-party \emph{qubit} states, it is well documented that the quantum state dimension has a large impact on the actual key rate \cite{Krenn:2017hz,Mirhosseini:2015fy,Groeblacher:2005ec,Mafu:2013jka,Lee:2016vk} and can significantly improve the robustness of such protocols against noise or other potential security leaks \cite{Huber:2013ex,Vertesi:2010bq}. Both of these properties make quantum key
distribution with $qudits$ a viable candidate for next-generation implementations. High-dimensional bipartite entanglement in the spatial and temporal degrees-of-freedom of a photon has been recently demonstrated in the laboratory \cite{Wang:2017ux,Martin:2017fk,Krenn:2014jy,Dada:2011dn,Jha:2008wu} and experimental methods for measuring high-dimensional quantum states are fast reaching maturity \cite{Bavaresco:2017wd, Islam:2017dd, Mirhosseini:2013em,Lavery:2012hi}. 

In parallel, recent years have seen the experimental realization of high-dimensional \emph{multipartite} entanglement \cite{Malik:2016gua, Erhard2017, Hiesmayr:2016eq}, as well as the development of techniques for generating a vast array of such states \cite{Krenn:2016ds}. These experimental advances signal that multi-partite high-dimensional entanglement is fast becoming experimentally accessible, thus paving the way for quantum communication protocols that take advantage of the full information-carrying potential of a photon.

The usefulness of multipartite entanglement for quantum key distribution was recently demonstrated
by designing QKD protocols which allow $n>2$ users to produce a secret key shared among all of them \cite{2016arXiv161205585E,Ribeiro2017}. Such a multipartite shared key can later be used, for example, for the secure broadcast of information. Both of these protocols use $n$-partite $GHZ$-type qubit states. In certain regimes, these protocols are  more efficient than sharing a secret key among $n$ parties via bipartite links followed by sharing of the broadcast key with the help of a one-time-pad cryptosystem. This advantage is especially pronounced in network architectures with bottlenecks (see \cite{2016arXiv161205585E}), making this protocol an interesting possibility for quantum network designs.
 
In this work, we go even further and generalize QKD schemes to protocols which use a general class of multipartite-entangled qudit states. Such states have an asymmetric entanglement structure, where the local dimension of each particle can have a different value \cite{Huber:2013ie,Huber:2013wo,Malik:2016gua}. The special structure of these states allows not only an increase in the information efficiency of the quantum key distribution protocol (either due to the dimension of the local states or the QKD network structure), but also adds a new qualitative property---multiple keys between arbitrary subsets or ``layers" of users can be shared simultaneously. Our generalization therefore shows a more complete picture of the advantages of multi-partite qudit entangled states in QKD networks, which goes beyond the simple increase in key rates. 


Let us now introduce the idea behind the proposed protocols with a simple motivating example. Consider a tripartite state 
\begin{equation}\label{eq:442}
\ket{\Psi_{442}} = \frac{1}{2}\left(\ket{000} + \ket{111} + \ket{220} +\ket{331}\right).
\end{equation}
After measuring many copies of this state locally in the computational basis, the three users---Alice, Bob and Carol---end up with data with interesting correlations. 
First of all, each of the four possible outcome combinations $000,111,220,331$
is distributed uniformly. Moreover, the outcomes of first two users (00, 11, 22, and 33) are perfectly 
correlated and partially independent of the outcomes of the third user. Alice and Bob can post-process their outcomes  into two uniform random bit-strings $k_{ABC}$ and $k_{AB}$ 
in the following way. 
\[
k_{ABC} =
\begin{cases}
  0 & \text{for outcomes } 0 \text{ and } 2 \\
1  & \text{otherwise}, 
\end{cases}
\]
and simultaneously 
\[
k_{AB} = 
\begin{cases}
0 & \text{for outcomes } 0 \text{ and } 1 \\
 1 & \text{otherwise}.
\end{cases}
\]
Note that $k_{ABC}$ is perfectly correlated to Carol's measurement outcomes, therefore it constitutes a random string shared between all three users. On the other hand, string $k_{AB}$ is completely 
independent of Carol's data---conditioned on either of Carol's two measurement outcomes, the value of $k_{AB}$ is $0$ or $1$, each with probability $\frac{1}{2}$. 
A simplified argument can now be made---since this procedure uses copies of pure entangled states, it is also independent of any other external data, therefore the strings $k_{ABC}$
and $k_{AB}$ are not only uniformly distributed, but also \emph{secure}.
It remains to show that this simple idea can be turned into a secure QKD protocol, in which a randomly chosen part of the rounds is used to assess the quality of the shared entanglement.

In Section \ref{sec1} we provide a protocol  which can implement an arbitrary layered key structure of $n$ users.
In Section \ref{sec2} we compare our proposed implementation
with the more conventional techniques of implementing key structures based on $EPR$ and $GHZ$-type states and show that aside from allowing very specific layered key structures, our proposed protocol provides a significant advantage in terms of key rates. In Section \ref{sec3} it is revealed that every layered key structure can be implemented with several different asymmetric multipartite high-dimensional states. Additionally, we study the relationship between local dimensions of the constructed states and the achievable key rates.

\section{Layered key structures and their implementation with asymmetric multipartite qudit states}\label{sec1}

\begin{figure*}[t]
\includegraphics[scale = 0.8]{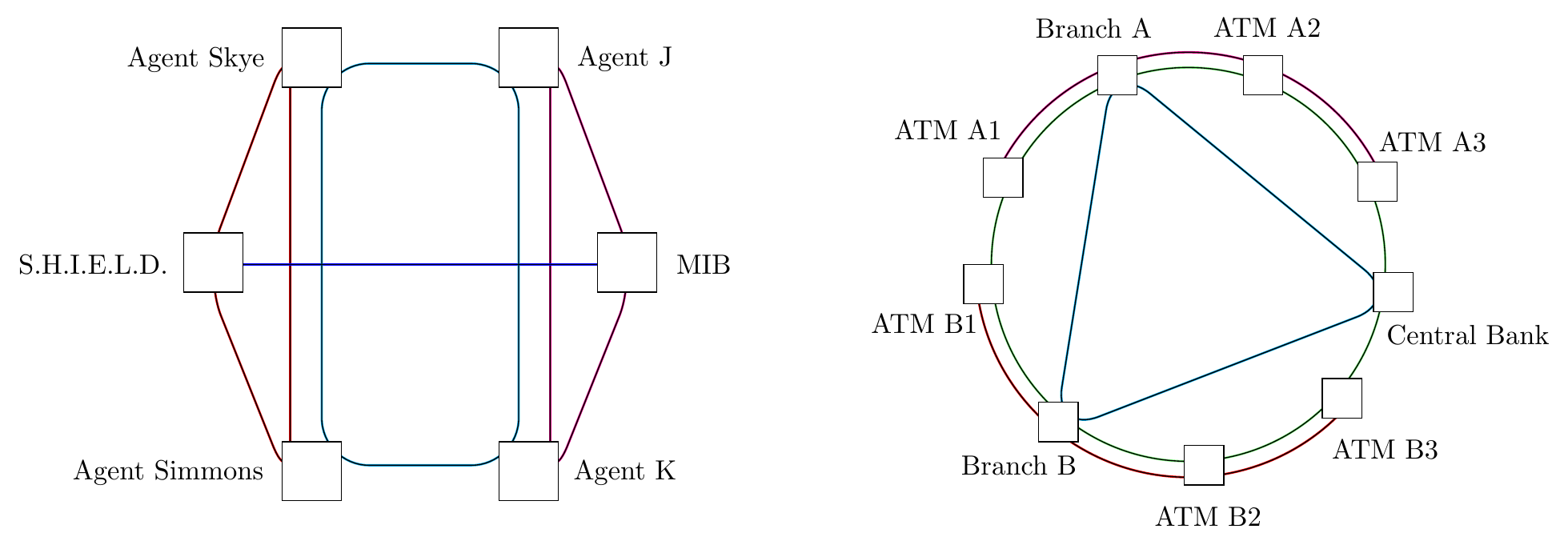}
\caption{\textit{Examples of layered key structures.} In the first example each of the two security agencies requires a secure communication channel with their respective agents, as well as an inter-agency channel secret even from their agents. Additionally, the agents require a secure channel shared only among themselves. In the second example, the central bank shares a key with each of its branches, while each branch shares a key with each of its ATMs and additionally, all parties share a common secret key.}
\label{fig:examples}
\end{figure*}

Suppose there are $n$ users of a quantum network. In order to achieve secure communication within this network, many types of shared keys are required. 
Apart from bipartite keys between pairs of users, which can be used for numerous cryptographic tasks such as encryption \cite{Vernam-Secretsignalingsystem-1919,Shannon-CommunicationTheoryof-1949} or authentication,
secret keys can be shared between larger groups of users. Also known as \emph{conference keys}, such keys have interesting uses such as secure broadcasting. Let us therefore define a \emph{layered key structure} as a set of keys required for secure communication in a given quantum network (see Figure \ref{fig:examples}).

Formally, we define a layered key structure $\mathcal{K}$ as a subset of the power set of users $\mathcal{K}\subseteq P\left(\mathcal{U}_n\right)$, where 
$\mathcal{U}_n$ denotes a set of $n$ users $\{u_1,\dots,u_n\}$. 
In order to conveniently talk about the layered key structures, let us define some of the parameters that describe them.
First, $K$ is the number of layers. Additionally, we will use the same labels for layers and keys shared in these layers. 
They are labeled by a natural number $i \in\{1,\dots, K\}$, therefore $i \in \mathcal{K}$ is a 
label for a single layer (key) of the layered structure. Last but not least, for each user $u_j$ let us define a parameter $\ell_j$ as the number of layers that the user $u_j$ belongs to, therefore $\ell_j:= \card{\{i\in \mathcal{K}\vert u_j \in i\}}$.

In what follows, given a particular key structure $\mathcal{K}$, we define a state that can be used for the implementation of $\mathcal{K}$ in a multipartite protocol.
The construction is based on implementations of correlations shared in a tensor product of $GHZ$ and $EPR$-type states for every layer with the help of high-dimensional states. 
\begin{algorithm}[H]
\floatname{algorithm}{State Preparation\\}
\renewcommand{\thealgorithm}{}
\caption{Given $\mathcal{K}$, find the state $\ket{\Psi_{\mathcal{K}}}$}
\label{protocol1}
\begin{algorithmic}[1]
\STATE For each layer $i$ that the user $u_j$ is part of, they hold a qubit labelled by $u^i_j$.
\STATE For each layer $i\in \mathcal{K}$, we define a state
\[
\ket{\psi_{i}} := \frac{1}{\sqrt{2}}
\left(\bigotimes_j\ket{0}_{u_j}+\bigotimes_j\ket{1}_{u_j}\right)
\]
\STATE Define the state 
              $\ket{\psi_{\mathcal{K}}} := \bigotimes_{i=1}^{K}\ket{\psi_i}$.
\STATE Each user $j$ encodes their $\ell_i$ qubits $\{u_j^i\}$ into a qudit 
register $u_j\pr$ of dimension $d_j = 2^{\ell_j}$ by rewriting binary string of qubits into digits.
\STATE The  resulting state $\ket{\Psi_\mathcal{K}}$ is an equal superposition of $2^K$ states of registers $d_1,\dots,d_K$.
\end{algorithmic}
\end{algorithm}

Before describing the QKD protocol for the layered key structure $\mathcal{K}$ implemented
with the state $\ket{\Psi_\mathcal{K}}$, let us first discuss the measurements we will use in the protocol. As stated above, each user $u_i$ holds a qudit state of dimension $2^{\ell_i}$. Our proposed protocol requires full projective measurements, therefore each user needs to be able to implement a projective measurement with $2^{\ell_i}$ outcomes. Additionally, since the state $\ket{\Psi_\mathcal{K}}$ can essentially be seen as a tensor product of various qubit $GHZ$ and $EPR$ states, the proof of security will be done by the reduction to multiple instances of protocols for such qubit states implemented simultaneously in higher dimensional systems. The protocols for qubit systems typically require only measurements in the three mutually unbiased qubit bases $\sigma_x,\sigma_y$ and $\sigma_z$ (see \cite{2016arXiv161205585E} for $GHZ$-based protocols and \cite{Renner2005} for an example of an $EPR$-based protocol). In order to use the analysis for a qubit state protocol for every layer, the user $u_j$ needs to implement measurements with $2^{\ell_j}$ outcomes that can be post-processed into measurement outcomes on the respective ``virtual'' qubits belonging to these layers. What is more, in order to keep the analysis of each layer independent, all combinations of qubit measurements are required. Let us therefore label required measurements of user $u_j$ as $M^j_{b_1,b_2,\dots,b_{\ell_j}}$, with $\forall i, b_i \in \{x,y,z\}$. Outcomes of such a measurement can be coarse-grained into measurement outcomes of measurements $\sigma_{b_i}$ on their respective qubits.

Let us now present the protocol:
\begin{algorithm}[H]
\floatname{algorithm}{The Layered QKD protocol\\}
\renewcommand{\thealgorithm}{}
\caption{Protocol for implementing $\mathcal{K}$ using $\ket{\Psi_{\mathcal{K}}}$}
\label{protocol2}
\begin{algorithmic}[1]
\STATE  In each round, user $u_j$ performs a randomly chosen projective measurement $M_{b_1,\dots,b_{\ell_j}}$ and coarse grains its outcome into measurement results for each ``virtual'' qubit corresponding to their layers.
\STATE The measurement choices are revealed to all users via a public channel.
\STATE For each layer $k_i$, the rounds in which $\sigma_z$ was measured by every user
in this layer are the key rounds.
\STATE  For each layer $k_i$, the rounds with other $\sigma_j$ measurement combinations are the test rounds.
\STATE In every layer separately, the test rounds are used for parameter estimation. 
\STATE Based on the parameter estimation results, error correction and privacy amplification are performed separately for every layer.
\end{algorithmic}
\end{algorithm} 

Note that this is truly a parallel implementation of the qubit protocols for all the layers using
higher dimensional qudit systems and it retains all the expected properties.
First of all, a particular round can be a key round for some of the layers and a test round for others. Moreover, it is possible, depending on the quality of the state, to have different key rates for each layer, including the situations when some of the layers have a key rate equal to $0$. And last but not least, the implementation and analysis of each layer $k_i$ does not depend on users who are not the part of this layer. In fact, each layer can be used and treated independently of the other layers. This signifies that the key in every layer $k_i$ is indeed secure even against other users, and additionally, it can be implemented even if the users of the network not in layer $k_i$ stop communicating.

\section{Comparison to other implementations of key structures}\label{sec2}
In this section we compare the performance of our protocol for implementing a key structure $\mathcal{K}$ with the performance of other possible implementations. 
The tools available for other implementations are the standard QKD protocols of two types:
\begin{enumerate}
\item Bipartite QKD protocols (qubit or qudit) for sharing a key between a pair of users with the use of EPR states such as
\[
\ket{\phi^+_d} = \frac{1}{\sqrt{d}}\sum_{i=0}^{d-1}
\ket{ii}.
\]
The qubit case of $d=2$ can be seen as the standard solution and is sufficient to implement any layered key structure with current technology. However, for the sake of a fair comparison we also allow for higher dimensional protocols 
(see e.g. \cite{Mirhosseini:2015fy}).
\item Recently, multi-party QKD protocols have been proposed that can implement a multipartite key with the use of GHZ-type states shared between $n$ users:
\[
\ket{GHZ_d^n}_{u_1,\dots,u_n} = \frac{1}{\sqrt{d}}\sum_{i=0}^{d-1}
\ket{ii\dots i}_{u_1,\dots,u_n}.
\]
Such protocols can be used to implement the key for each layer separately. Although so far only qubit $(d=2)$ protocols are known \cite{2016arXiv161205585E,Ribeiro2017}, we also allow for protocols with higher dimensional systems, which are in principle possible.
\end{enumerate}

These existing protocols can be combined to implement the given layered key structure $\mathcal{K}$ in multiple ways. Here we compare the performance of two specific implementations. The first one uses only bipartite QKD protocols of various dimensions between the selected pairs of users. These bipartite keys are subsequently used to distribute a locally generated multipartite key via one-time-pad encryption \cite{Vernam-Secretsignalingsystem-1919}. The second implementation uses the $GHZ$ protocols of various dimensions to directly distribute the keys for each layer. 

The merit of interest is the \emph{idealized key rate} $r_i$ associated with every 
layer $k_i$. 
The idealized rate $r_i$ is the expected number of key bits in the layer $k_i$ per the time slot, under an assumption that only key round measurements (i.e. the computational basis) are used.
Such a merit captures how efficiently the information carrying potential of the photon is used
in different implementations, neglecting the need for the test rounds used in the parameter
estimation part of the protocol.

In order to further specify what implementations of the layered key structure $\mathcal{K}$
we are comparing to, we need to characterize two different properties of the quantum network we are using for comparison.

Since the achievable idealized rates depend on the architecture of the network
(as illustrated in \cite{2016arXiv161205585E}), let us specify the network architecture first. Let us suppose that the $n$ users $\mathcal{U}_n$ form a network where each $u_i$ is connected to a source of entanglement by a quantum channel, and each pair of users $(u_i,u_j)$ shares an authenticated classical channel (see Figure \ref{fig:entDistr}).

\begin{figure}\label{fig:entDistr}
\begin{center}
\includegraphics[scale = 1.5]{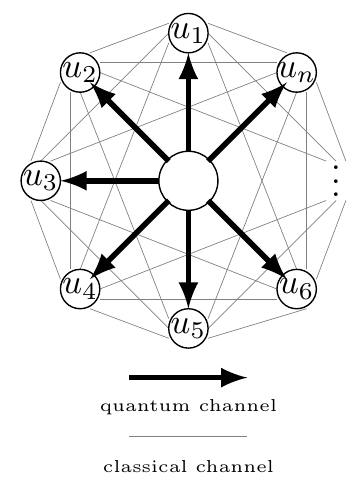}
\caption{\textit{Entanglement distribution model.} Each user $u_j$ is connected to the source of entanglement via a quantum channel. Additionally each pair of users shares a classical channel.}
\label{fig:entDistr}
\end{center}
\end{figure}

The second property of the network we need to specify are the local dimensions of the measurements allowed for each user. We restrict every user
to the local dimension of $\ket{\Psi_{\mathcal{K}}}$ -- user $u_i$ can perform projective measurements with at most $2^{\ell_i}$ outcomes. This is a reasonable assumption, since it is a statement about the complexity of the measurement apparatus of each user $u_i$.
This choice of dimension is also meaningful, since 
in a certain sense, our protocol is a good benchmark implementation under these local dimension assumptions. It achieves the rates $r_i =1$ for all layers $i$ and it is
not difficult to see that this is impossible with lower local dimensions, since the logarithm
of the local dimension $d_i$ of the user $u_i$ needs to be at least $\ell_i$ -- the number of shared bits in each round.

Note that the two aforementioned assumptions do not restrict the routing capabilities of the source. This means that the source can send out entangled states to any subset of users on demand. Also, these assumptions allow for simultaneous distribution of entangled states to mutually exclusive sets of users. Therefore, for example, in networks of $2n$ users, $n$ $EPR$ pairs can be sent simultaneously, or,  alternatively two $n$-partite $GHZ$ states can be sent simultaneously and so on. The routing capabilities required of a source 
in order to be able to implement such approaches pose significant experimental challenges---for example, in access QKD networks 
\cite{ChangDengYuanEtAl-Experimentalrealizationof-2016,ChoiYoungTownsend-Quantuminformationto-2011,Frohlich2013} 
only a single pair of users can receive
an $EPR$ pair in a single time slot. However, for the sake of a fair comparison we allow them anyway. 
Note that in this sense our protocol is passive, since the source produces the same state in every round of the protocol.

\begin{figure}[t]
\begin{center}
\includegraphics[scale = .5]{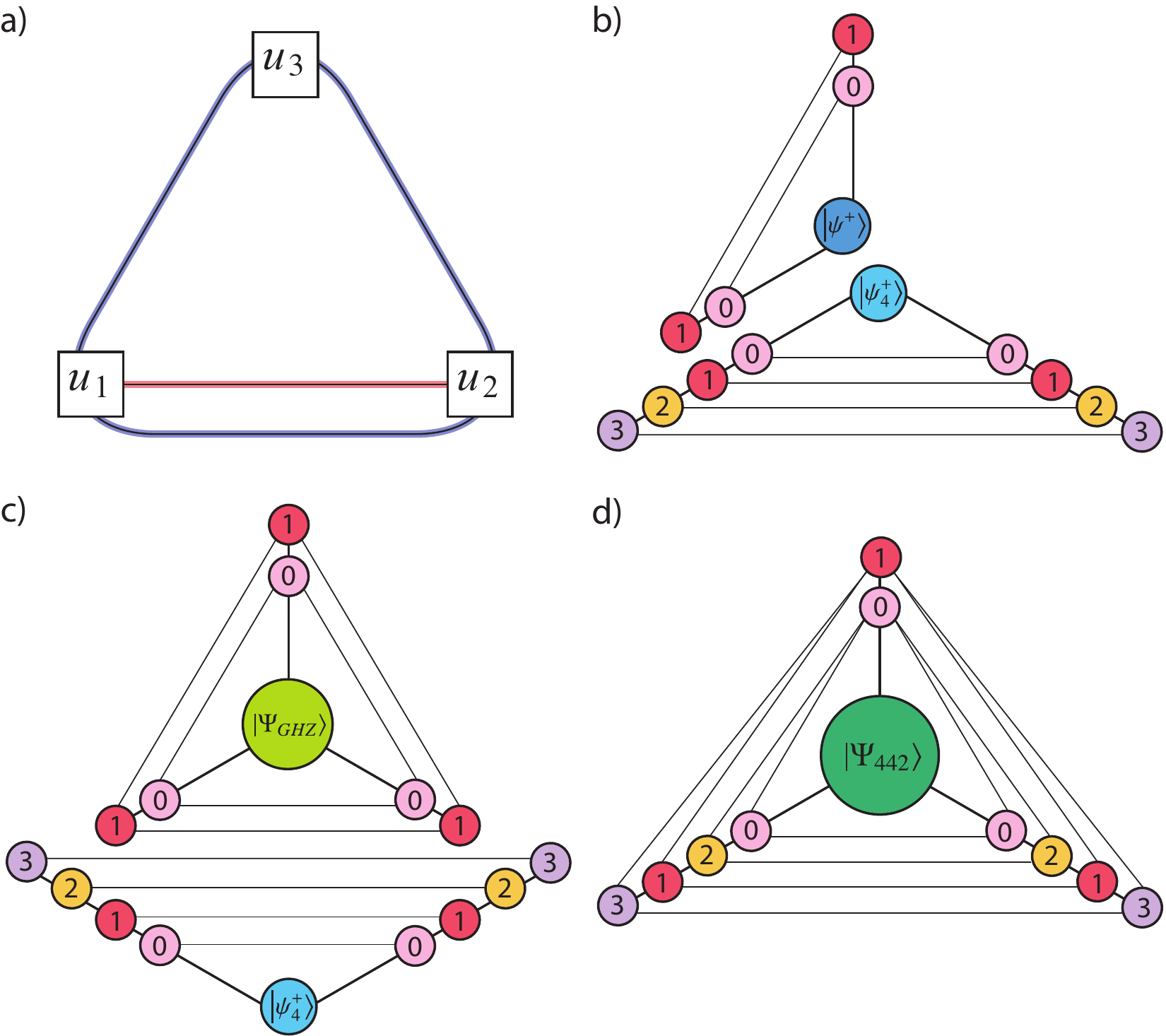}
\caption{\textit{Simplest LQKD example.} a) In this example, three users want to share keys in two layers $k_1 = \{1,2,3\}$ and $k_2 = \{1,2\}$. $p$ is the probability that users 1 and 2 share the state $\ket{\psi_4^+}$\\
b) An EPR implementation results in the idealized rates $[\{1,2,3\};(1-p)],[\{1,2\};3p-1]$.\\
c) A GHZ implementation results in the idealized rates $[\{1,2,3\};(1-p)],[\{1,2\};2p]$.\\
d) An implementation with the state $|\Psi_{442}\rangle=\frac{1}{2}(|000\rangle+|111\rangle+|220\rangle+|331\rangle)$ results in the idealized
rates $[\{1,2,3\};1],[\{1,2\};1]$.}
\label{fig:simplestLQKD}
\end{center}
\end{figure}

In order to familiarize the reader with our setup, we explicitly calculate the idealized rates for the simplest case of three users (Alice ($1$), Bob ($2$) and Carol ($3$)), with the layered key structure $\{k_1 = \{1,2,3\},k_2 = \{1,2\}\}$ (see Fig.~\ref{fig:simplestLQKD}a), before discussing the rates of different implementations more generally. First of all, for this layered key structure $\mathcal{K}$, the associated state is the one introduced in Section I:
\[
\ket{\Psi_{442}} = \frac{1}{4}
\left( 
\ket{000} +\ket{111} +\ket{220} + \ket{331}
\right).
\]
This fixes the local dimensions to $4$ for Alice and Bob and $2$ for Carol.

Furthermore, note that in a network of just three users, an EPR pair can be sent only to a single pair of users in each time slot. However, since Alice and Bob can perform ququart measurements, they can use any given time slot to share and run a ququart QKD protocol with the state $\ket{\psi^+_4} = \frac{1}{4}\left(\ket{00} + \ket{11} + \ket{22} +\ket{33} \right)$,
achieving the idealized rate of $2$.

Therefore, in order to implement the given key structure, the source will alternate between sending an $EPR$ pair $\ket{\psi^+_4}$ to Alice and Bob with probability $p$, and sending a standard qubit (since Carol can manipulate only qubits) EPR pair $\ket{\psi^+}$ to Alice and Carol with probability $(1-p)$ (see Fig.~\ref{fig:simplestLQKD}b). This results in an idealized rate  $r_{AB} = 2p$ for the bipartite key $k_{AB}$ between Alice and Bob. The rate of the key $k_{AC}$ between Alice and Carol in this setting is  $r_{AC} = (1-p)$. In order to get one bit of the desired key $k_{ABC}$, a bit of each key $k_{AB}$ and $k_{AC}$ needs to be used---Alice locally generates a secret string
$k_{ABC}$ and sends an encrypted copy to both Bob and Carol. Therefore, exchanging \emph{all} bits of key $k_{AC}$ and an equivalent amount of key $k_{AB}$ in this way results in the rates $[\{1,2,3\};(1-p)],[\{1,2\};2p-(1-p)]$. Note also that values of $p\leq\frac{1}{3}$ do not allow the users to exchange all keys $k_{AC}$ into tripartite keys, since the amount of the keys $k_{AB}$ is too low. For comparison, note that our layered implementation results in the rate $[\{1,2,3\};1],[\{1,2\};1]$, while the previous analysis suggests that keeping the rate $r_{AB} = 1$ ($p=\frac{2}{3}$) results in  $r_{ABC} = \frac{1}{3}$.

The analysis for the $GHZ$ implementation is much more simple. Here, either the source sends a qubit $GHZ$ state with probability $(1-p)$, or a ququart $EPR$ state to Alice and Bob with probability $p$ (see Fig.~\ref{fig:simplestLQKD}c). This results in the rates $[\{1,2,3\};(1-p)],[\{1,2\};2p]$. For comparison, keeping the rate  $r_{AB} = 1$ ($p=\frac{1}{2}$) results in  $r_{ABC} = \frac{1}{2}$. Thus, while this implementation is more efficient than the $EPR$ one, it still cannot achieve the rate of $1$ for both layers obtained by the state $\ket{\Psi_{442}}$ shown in Fig.~\ref{fig:simplestLQKD}d.

The problem of finding the general form of achievable rates for an arbitrary key structure $\mathcal{K}$ is too complex and would involve too many parameters.
The reason for this is the fact that the probabilities (or in fact \emph{ratios}) of $EPR$ or $GHZ$ states sent to the different
subsets of users change the average rates $r_i$ in different layers (see the previous example).
Therefore the goal of the following subsection is to argue that the rates $r_i = 1$ for all $i$ are achievable for only restricted classes of key structures $\mathcal{K}$ with both $EPR$ 
and $GHZ$ implementations. 

\subsection{Connected structures and partitions}

Naturally, each layered structure $\mathcal{K}$ defines a neighborhood graph  $G_{\mathcal{K}}$. Users  $\mathcal{U}_n$ are represented as the vertices in this graph
and two users $u_i$ and $u_j$ are connected by an edge, if they share a layer in the
structure $\mathcal{K}$. We call a layered structure $\mathcal{K}$ connected, if the 
neighborhood graph $G_{\mathcal{K}}$ associated to it is connected.

 The connected components of each layered structure $\mathcal{K}$ can be 
treated separately, since the source can send states to them simultaneously and therefore their rates do not depend on the rates of the other connected components.
In what follows, we therefore deal only with connected key structures $\mathcal{K}$.

\begin{figure}[b]
\begin{center}
\includegraphics[scale = 1.2]{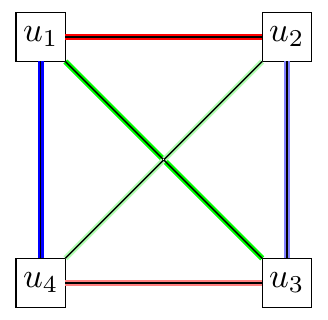}
\caption{\emph{A key structure with idealized rate $1$ with $EPR$ implementation.}
There are $6$ two-user layers, which can be grouped into $3$ partitions (red, green and blue). $8$-dimensional $EPR$ pairs can be
distributed to each partition in parallel. If each of these distribution rounds happens with 
probability $\frac{1}{3}$, the average rate for every layer is $1$.}
\label{fig:EPRrate1}
\end{center}
\end{figure}

Let us now introduce \emph{partitions $P_i$} of the key structure $\mathcal{K}$. 
These are
subsets of layers that are mutually exclusive and collectively exhaustive -- meaning that
their union is equal to the set of all users $\mathcal{U}_n$ and no pair of the layers in the partition contain the same user. Formally:
\[
\mathcal{P}_i = \left\{k^i_1,\dots,k^i_m\vert \cup_j k^i_j = \mathcal{U}_n, \forall a,b: k^i_a \cap k^i_b = \emptyset \right\}.
\]
Note that we maintain an index $i$ for each partition, since each connected 
layered structure might contain several partitions (see Figure \ref{fig:EPRrate1}).

Let us now suppose that all the layers of a key structure $\mathcal{K}$ can be grouped 
into exactly $\ell$ partitions.
In such a case, each user belongs to exactly $\ell$ layers and therefore 
$\forall i: \ell_i = \ell$.
We will show that for the $GHZ$ implementation, all the partitions $\mathcal{K}$ with this property
can achieve the idealized rate $r_i = 1$ for all layers. For the $EPR$ implementation to achieve all rates equal to $1$, an additional requirement is needed---all layers need to
have size, given by the number of users, of $2$.

The crucial observation is that the source can send a $GHZ$ state of dimension $2^\ell$ to
each layer in a partition $\mathcal{P}_i$ simultaneously, resulting in rate $\ell$ in each of these layers.
It takes the source exactly $\ell$ time slots to iterate over all the partitions $\mathcal{P}_i$,
therefore the average rate for each layer is $1$. For the case of all the layers being of
size $2$, this simple distribution protocol reduces to one with $EPR$ pairs.

It remains to be shown that key rates of $1$ cannot be achieved in every layer, unless the key structures $\mathcal{K}$ can be grouped into
partitions of $\mathcal{U}_n$ without leftover layers. To see this, it is enough to carefully count the number of key bits that are required to be produced in every time step. In order to achieve the rate $1$ in each layer, each user needs
to produce a total of $\ell_i$ secret key bits in every round. This can only be achieved if every user measures a state of full dimension in every
time step. However, this is not possible for connected key structures $\mathcal{K}$ that cannot
be fully decomposed into multiple partitions. To see this, consider a user $u_i$. In order to realize the full information-carrying potential, the user $u_i$
needs to share a $2^{\ell_i}$-dimensional $GHZ$ state in one of his layers in a single round. This implies that all the neighbors $\{u_j\}$ of user $u_i$ have $\ell_j = \ell_i$, since otherwise they
either won't be able to measure in $2^{\ell_i}$ dimensions, or they will not be able to generate enough key in the given round. This fact, together with the connectedness of
the key structure, implies that $\ell_j = \ell_i$ for all users. In the case of $\ell_i = 1$, the desired graph is not connected. Let us therefore discuss only key structures with $\ell_i>1$.
In each round, each user needs to share a key in one of his layers. This is possible only if each layer is a part of a partition. Additionally, since each user has $\ell_j = \ell_i$,
to obtain the rate $1$ in every layer, each user needs to iterate over all his layers in exactly $\ell_i$ rounds. This implies that the key structure can be decomposed into $\ell_i$ partitions.

\begin{figure}[h]
\begin{center}
\includegraphics[scale = 1.5]{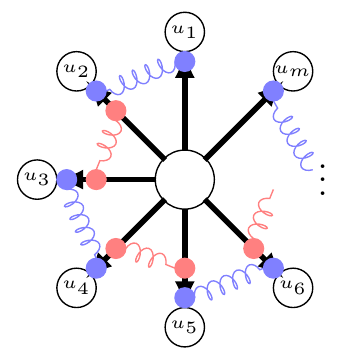}
\caption{\textit{The number of channel uses needed in order to share a multipartite key with $EPR$ pairs.} Users $u_1$ and $u_m$ need to share only a single $EPR$ pair with their neighbors. The rest of the users need to share
two $EPR$ pairs each. To share the multipartite key, user $u_1$ generates a random string locally and sends it to the user $u_2$ secretly via one time pad encryption. Then each user $u_i$, 
after receiving the key from the user $u_{i-1}$, sends it secretly to the user $u_{i+1}$, until all the users share the new secret key.}
\label{fig:EPRimplementation}
\end{center}
\end{figure}

An $EPR$ implementation requires an additional restriction on the key structures implementable with rate $1$.
The reason for this fact is that in each layer of size $m>2$, there is a user
who needs to generate two bits of bipartite key in order to securely distribute
the locally generated multipartite key (see Figure \ref{fig:EPRimplementation}). The number of required bits per round therefore exceeds $\ell_i$ in some rounds for some of the users, whenever
there is a key shared among a number of users larger than $2$. This fact shows that even if the key structure can be grouped into partitions, with all users having the same local dimension $\ell$ and generating $\ell$ bits of bipartite randomness in each round, there are some users who need to generate more than $\ell$ bipartite key bits in order to share $\ell_i$ bits in their multipartite layers. In fact, this additional requirement has a very simple corollary---if the number of users $n$ is even, the $EPR$ implementation cannot achieve the idealized rate $1$ in each layer.

\section{Dimension-rate trade-off}\label{sec3}
In this section we show how to construct many different multipartite high-dimensional states that are useful for the implementation of a given key structure $\K$. 
These states differ from each other in their local dimensions as well as achievable
idealized key rates---generally there is a trade-off between these two quantities.

As an example, consider the layered structure $\mathcal{K} =\left\{\{1,2,3\},\{1,2\}\right\}$ depicted in Figure \ref{fig:simplestLQKD}. The solution discussed previously
can be used to implement this layered structure with the $\ket{\Psi_{442}}$ state (see eq. \eqref{eq:442}) of local dimensions $4$ for the first two users and $2$ for the third user. However, consider the following state:
\begin{align}\label{eq:332}
|\Psi_{332}\rangle=\frac{1}{\sqrt{2}}\left(|000\rangle+\frac{1}{\sqrt{2}}(|111\rangle+|221\rangle)\right)\,,
\end{align}
which is very close to the first such asymmetric state that was recently realized in the lab \cite{Malik:2016gua}. Measuring the state in the computational basis produces data that can be post-processed into two uniformly random and independent keys in the
following way:
\[
k_{123} =
\begin{cases}
0 & \text{for outcome } 0 \\
1  & \text{otherwise}, 
\end{cases}
\]
while simultaneously 
\[
k_{12} = 
\begin{cases}
0 & \text{for outcome } 1\\
1 & \text{for outcome } 2\\
\perp & \text{otherwise,}
\end{cases}
\]
where $\perp$ denotes that no key was produced in this layer. The idealized rate associated with this state is 
$[\{1,2,3\};1],\left[\{1,2\};\frac{1}{2}\right]$, as
a bit for the key $k_{1,2}$ gets produced only with probability $\frac{1}{2}$. Interestingly, a comparison with other implementations (see Figure 
\ref{fig:simplestLQKD}) reveals that even though the local dimensions of the $\ket{\Psi_{332}}$ state are more restricted, it can nonetheless achieve rates that are unattainable by four-dimensional implementations for separate layers.
In this section, we discuss under which conditions such local dimension-rate trade-off is possible and subsequently use this knowledge in order to construct a whole family of states which are useful for the implementation of a given layered key structure $\K$.

The main idea allowing for the dimension-rate trade-off is not to produce key bits in some of the layers for certain 
measurement outcomes, which results in a smaller local dimension.
However, this idea is not usable in every situation, since even the measurement outcome post-processed to $\perp$ can leak 
information about the key produced in different layers.
In order to show this, consider two layers $k_i$ and $k_j$ with $k_i \cap k_j \neq \emptyset$. Additionally, label the user 
present in both layers as $u$. Without loss of generality, assume that user $u$ interprets the measurement outcome $a$ as $\perp$ in 
layer $k_i$ and as a key bit $0$ in layer $k_j$. Since all the users in the layer $k_j$ are fully correlated, the key shared in this 
layer can be interpreted as a string of symbols 
$0,1,$ and $\perp$. It is important to notice that keys in layers $k_i$ and $k_j$ are not independent. While the users of layer 
$k_i$ can infer only that no key was produced in layer $k_j$ in rounds where a bit $0$ was produced in $k_i$ (which is not a 
security breach), users of $k_j$ know that whenever the protocol produced no key symbol $\perp$ in layer $k_j$, a bit $0$  
was produced in layer $k_i$. 
This is not a security breach if and only if all users of the layer $k_j$
are authorized to also know the key $k_i$, in other words, if and only if $k_j\subset k_i$.

In order to explain how to use this observation in the construction of states
for any layered structure $\K$,
let us first revisit the state construction algorithm proposed in section \ref{sec2}
and reformulate it recursively. Consider two layered key structures $\mathcal{K}_1$ with users $\U_1$ 
implementable with a state $\ket{\Psi_{\K_1}}$ and 
$\mathcal{K}_2$ with users $\U_2$ 
implementable with a state $\ket{\Psi_{\K_1}}$.
A new layered key structure $\K := \K_1 \cup \K_2$ with users $\U = \U_1 \cup \U_2$ 
can be implemented with a state $\ket{\Psi_{\K}}$, constructed as follows:
\begin{algorithm}[H]
\floatname{algorithm}{Recursive State Preparation step 1\\}
\renewcommand{\thealgorithm}{}
\caption{Given $\ket{\Psi_{\K_1}}$ and $\ket{\Psi_{\K_2}}$, find the state $\ket{\Psi_{\mathcal{K}}}$ \\with $\K = \K_1\cup\K_2$}
\label{union1}
\begin{algorithmic}[1]
\STATE Consider the state 
              $\ket{\psi_{\mathcal{K}}} := \ket{\Psi_{\K_1}} \otimes \ket{\Psi_{\K_2}}$.
\STATE Each user $u_j\in \K_1 \cup \K_2$ holds two registers $u_j^1$  and $u_j^2$ of dimensions ${d^1_j}$ and 
${d^2_j}$ respectively.
\STATE Let each user $u_j \in \K_1 \cup \K_2$ encode their registers $u_j^1$ and $u_j^2$ in
a register $u\pr_j$ of dimension $d\pr_j = d_j^1d_j^2$.
\STATE The  resulting state is the desired state $\ket{\Psi_\mathcal{K}}$.
\end{algorithmic}
\end{algorithm} 

The local dimensions of the resulting state are  $d\pr_j = d_j^1d_j^2$ for each user $u_j \in \U_1\cap \U_2$ and remains unchanged
(\textit{i.e.} $d\pr_j = d_j$) for all users $u_j \notin \U_1\cap \U_2$.
Consider a layered key structure $\K$ with $\vert\K\vert$ layers. If we assign a qubit $k$-partite $GHZ$ state to each of the layers of size $k$, we can 
recover the state for $\K$ constructed in Section \ref{sec1} by simply joining the $GHZ$ states one by one with the recursive step 1 we just introduced. 


In order to incorporate the dimension-rate trade-off into the state construction, let us present an alternative recursive step that takes two states $\ket{\Psi_{\K_1}}$ and $\ket{\Psi_{\K_2}}$ as an input. These two states implement key 
structures $\K_1$ and $\K_2$ with users $\U_1$ and $\U_2$ respectively. The recursive step produces
a state $\ket{\Psi_{\K}}$, which implements key structure $\K = \K_1 \cup \K_2 \cup \left\{\U_1 \cup \U_2\right\}$,
where $\left\{\U_1 \cup \U_2\right\}$ is a new layer containing all users in both $\U_1$ and $\U_2$.

\begin{algorithm}[H]
\floatname{algorithm}{Recursive State Preparation step 2\\}
\renewcommand{\thealgorithm}{}
\caption{Given $\ket{\Psi_{\K_1}}$ and $\ket{\Psi_{\K_2}}$, find the state $\ket{\Psi_{\mathcal{K}}}$ \\with $\K = \K_1 \cup \K_2 \cup \left\{\U_1 \cup \U_2\right\}$}
\begin{algorithmic}[1]\label{alg3step1}
\STATE Consider a state $\ket{\Psi_{\K_2}\pr}$, which is equal to $\ket{\Psi_{\K_2}}$, but with all labels of computational basis vectors primed.
\STATE A state implementing $\K$ can be written as:
\begin{align*}
\ket{\Psi_\K} &:= \frac{1}{\sqrt{2}} \left(\ket{\Psi_{\K_1}}_{\U_1} \otimes \ket{\perp,\dots,\perp}_{\U_2\backslash\U_1}\right)\\
&+ \frac{1}{\sqrt{2}} \left(\ket{\Psi\pr_{\K_2}}_{\U_2} \otimes \ket{\perp,\dots,\perp}_{\U_1\backslash\U_2}\right),
\end{align*}
where $\perp$ is a new symbol.
\end{algorithmic}
\end{algorithm} 

The local dimensions of state $\ket{\Psi_{\K}}$ are $d_j^1 + d_j^2$ for users $u_j\in \U_1 \cap \U_2$ and $d_j+1$ for users
$u_j \notin \U_1\cap \U_2$. The reason for this is that in the construction we are using primed labels of the computational basis states
of $\ket{\Psi_{\K_2}}$ together with the original basis labels of the state  $\ket{\Psi_{\K_1}}$. The resulting states of users
$u_j\in \U_1 \cap \U_2$ therefore effectively live in a Hilbert space obtained by a direct sum of their original Hilbert spaces. The addition of one
dimension for the remaining users comes from the fact that we enlarge their computational basis with a new symbol $\perp$ in the 
construction.

Note that neither $\K_1$ and $\K_2$ are necessarily non-empty in the construction. For this reason, let us define a state for $\K =\emptyset$ with $n$ users
as $\ket{\Psi_\emptyset} = \ket{00\dots 0}_{u_1,\dots,u_n}$. This is especially important in order to be able to use the trade-off recursive step
to construct a state for a union of two key structures $K_1$ and $K_2$, such that $\U_1 \subseteq \U_2$ and $\K_2 = \U_2$, i.e.~the layered key structure
$\K_2$ contains only a single layer---the set of all of its users (see Figure \ref{fig:simplestLQKD} for an example of such a key structure). In such a case, a state for 
the implementation of $\K = \K_1 \cup \{\U_2\}$ can be constructed with the recursive step $2$ applied to states $\ket{\Psi_{\K_1}}_{\U_1}$ and 
$\ket{\Psi_{\emptyset}}_{\U_2}$.

Now we would like to discuss how to use the state $\ket{\Psi_{\K}}$ to construct a QKD protocol producing a key in all the layers of the
key structure $\K$. Our argument is again structured along the lines of a reduction to existing qubit QKD protocols for $GHZ$ states of $n$ users. The key observation for 
a state $\ket{\Psi_\K}$ with user set $\U = \U_1\cup\U_2$ created with the recursive step $2$ is that it is an equal superposition of two computational basis
vectors, which are not only orthogonal, but also differ in every position. We call this property 
\textit{local distinguishability}. Let us now divide the Hilbert spaces of the users in $\U$ into two 
orthogonal components.
Users $u_j\in \U_1\backslash \U_2$ can split their Hilbert space into two orthogonal subspaces spanned by $\left\{\vspace{-2mm}\{\ket{i}\}_{i=1}^{d_j^1},\ket{\perp}\right\}\in\Hilb_{d_j^1}\oplus\Hilb_1$, 
where the set of orthogonal computational basis
vectors $\{\ket{i}\}_{i=1}^{d_j^1}$ is the computational basis of the Hilbert space of user $u_j$ in the input state $\ket{\Psi_{\K_1}}$. Similarly, users 
$u_j \in \U_2\backslash \U_1$ can split their Hilbert spaces into two orthogonal subspaces spanned by $\left\{\ket{\vspace{-2mm}\perp},\{\ket{i\pr}\}_{i=1}^{d_j^2}\right\}\in\Hilb_{1}\oplus\Hilb_{d_j^2}$, where
$\{\ket{i\pr}\}_{i=1}^{d_j^2}$ is the orthogonal computational basis of the Hilbert space of user $u_j$ in the input state $\ket{\Psi_{\K_2}}$. Finally,
users $u_j\in \U_1\cap\U_2$ can split their Hilbert spaces into two orthogonal subspaces spanned by $\left\{\{\ket{i}\}_{i=1}^{d_j^1},\{\ket{i\pr}_{i=1}^{d_j^2}\}\right\}\in\Hilb_{d_j^1}\oplus\Hilb_{d_j^2}$,
where $d_j^1$ and $d_j^2$ are the dimensions of the Hilbert spaces of user $u_j$ in the input states $\ket{\Psi_{\K_1}}$ and $\ket{\Psi_{\K_2}}$ respectively.
For each user, projectors onto these two subspaces define an incomplete set of POVM measurements, which can be used as analogues of $\sigma_z$ measurements in the key rounds of the QKD protocol.
Since these states are fully correlated in the respective subspaces, there are two kinds of possible classical global measurement outcomes, each
occurring with probability $\frac{1}{2}$. Either the outcome is $(i,i,\perp)$ or $(\perp,i\pr,i\pr)$. By a simple renaming the first outcome can be seen
as a $0$ shared among all the users and the second one as $1$, thus constituting a common shared binary key. However, since the measurements (depending on the
dimensions of $\ket{\Psi_{\K_1}}$ and $\ket{\Psi_{\K_2}}$ do not have to be fully informative, they also lead to interesting post-measurement states. The post measurement
states are respectively $\ket{\Psi_{\K_1}}_{\U_1}\otimes\ket{\perp\dots\perp}_{\U_2\backslash\U_1}$ and $\ket{\perp\dots\perp}_{\U_1\backslash\U_2}\otimes\ket{\Psi_{\K_2}}_{\U_2}$.
Clearly the post measurement states can be used to implement sub-key structures $\K_1$ and $\K_2$ by their respective users.

The analogues of $\sigma_y$ and $\sigma_z$ needed to formulate the full GHZ protocol \cite{2016arXiv161205585E} require first projecting
the state $\ket{\Psi_{\K}}$ down to a qubit state by locally mapping all vectors in the left Hilbert space of each user onto a state $\ket{0}$ and all vectors in the right
Hilbert space onto a state $\ket{1}$, followed by qubit measurements $\sigma_x$ and $\sigma_y$. The down-projection results in a loss of information about the exact position 
of the vectors in their respective subspaces. However, this happens
only in the test rounds in which we cannot use the post-measurement states to implement the keys in the sub-structures anyway. In other words, each layer is probed only probabilistically in the parameter estimation rounds. However, since the parameter estimation rounds are generically sub-linear, this only leads to a constant increase of a sub-linear number of rounds and does not impact the key rounds at all.
Since only a small (logarithmic) portion of test rounds is required, most of the
states will be measured in $\sigma_z$ measurements. Post-measurement states of these measurements will be useful for the implementation of $\K_1$ half the time on average, 
and the other half will be useful for the implementation of $\K_2$. This probabilistic nature of obtaining the post-measurement states is the source of the rate decrease in this 
construction. Repeating the reductions to binary QKD protocols for the sub-states leads to recovering a QKD protocol for the key structure $\K$.

A state for any layered key structure $\K$ can be constructed by starting with an empty layered key structure
and subsequent application of one of the previous recursive rules until all the layers of $\K$ have been added.
It is important to note that the exact form of the resulting state---\textit{i.e.} the local dimensions and the idealized rates in every layer---depends on the types of recursive steps
we use for each layer, but also on the order. This is because adding the layers of the structure $\K$ in particular orders might result in the inability to use the trade-off rule.

The simplest example to consider is once again the key structure
$\K = \{\{1,2\},\{1,2,3\}\}$. Note that the recursive rule number $1$ can only join two non-empty key structures into one state. In principle, this is not a problem, since we know that for layered key structures with only a single layer of size $n$, there is only a single suitable state---the GHZ state of $n$ users.  
Therefore, we can start by dividing $\K$ into the single layer sub-layers $\K_1 = \{1,2,3\}$ and $\K_2 = \{1,2\}$ and assigning to them their respective binary GHZ and EPR states.
After doing this, we cannot use the recursive rule number $2$ anymore. Therefore, the only option is to join the states together with the recursive rule number $1$ resulting in the
$\ket{\Psi_{442}}$ state \eqref{eq:442}.

Another option is to assign an EPR pair to the layer $\{1,2\}$ and subsequently use the second recursive rule
with $\K_1 = \{1,2\}$ and $\K_2 = \{\emptyset\}$, with user sets $\U_1 = \K_1$ and 
$\U_2 = \{1,2,3\}$ implemented with states
$\ket{\Psi_{\K_1}} = \frac{1}{\sqrt{2}}\left(\ket{00} + \ket{11}\right)$ and $\ket{\Psi_{\K_2}} = \ket{000}$,
in order to obtain state 
\begin{align}
\ket{\Psi_\K} = \frac{1}{\sqrt{2}}\left(\frac{1}{\sqrt{2}} \left(\ket{00\perp} + \ket{11\perp}\right) + \ket{0\pr0\pr 0}  \right),
\end{align}
which is equivalent to the state $\ket{\Psi_{332}}$ defined in Eq.~\eqref{eq:332}.

\begin{figure}
\includegraphics[scale = 0.7]{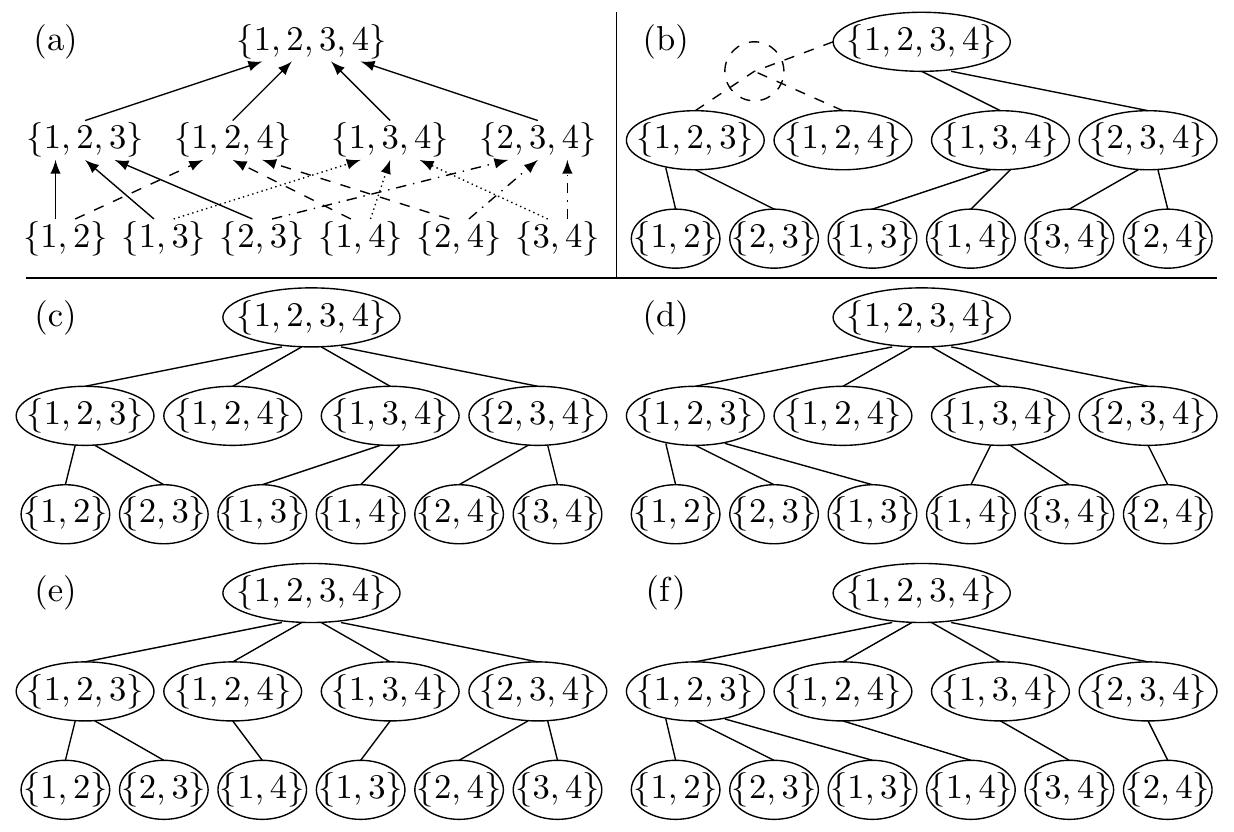}
\caption{\textbf{Classification of different states for key structure $\K= P(\{1,2,3,4\})\backslash\{\emptyset,\{1\},\{2\},\{3\},\{4\}\}$.} 
(a) Key structure $\K$ ordered with respect to inclusion. 
(b) Decomposition of ordered $\K$ into binary trees. Note that union of two children vertices is always equal to their parent.
Each tree can now be implemented using the trade-off recursive step multiple times. Joining the trees together can
be done by the recursive step $1$. This is illustrated by the dotted circle that joins three states together. (c)-(f) Different decompositions of ordered $\K$ into a non-binary tree. Using reductions to qudit protocols allows us to use the trade-off 
recursive step for all layers. Each decomposition leads to a different state.}
\label{fig:PSTD}
\end{figure}

Let us study the family of states for a fixed key structure $\K$ in more detail. Recursive rule number $2$ can be used to join two 
key sub-structures  $\K_1$ and $\K_2$ only if the final key structure $\K$ also contains a layer $\U_1 \cup \U_2$. For this reason, 
the central concept of this part of the section is ordering the layers of the key structure $\K$ with respect to the set inclusion (see Figure \ref{fig:PSTD} (a)).

The first step necessary to characterize different states that can be prepared for a given $\K$ using the introduced recursive rules is to first order the layers $k\in\K$ according to the inclusion. This ordering can be represented by an ordered graph $\tilde{G}_\K$, where each layer  is represented by a vertex
and two vertices are connected if and only if one is a subset of the other (see Figure \ref{fig:PSTD} (a)).
The next step is to find specific binary \emph{tree decompositions} of $\tilde{G}_\K$. 

A tree decomposition is a division of the graph into tree subgraphs, where all the vertices are used 
and the tree subgraphs are connected by edges from the edge set of graph $\tilde{G}_\K$. An additional condition for the decompositions suitable for our purposes is that
the trees should correspond to key sub-structures which can be implemented with the help of recursive rule $2$ only. 
This condition translates to the fact that the trees in the decomposition have to fulfill an additional constraint---the union of two children vertices has to be equal to 
their parent vertex (see Figure \ref{fig:PSTD} (b)).

By construction, each of the key sub-structures corresponding to a tree can be implemented using the recursive rule $2$ only by assigning qubit $GHZ$ states with the correct number of parties to the layers corresponding to leaves in the tree. Following this, the state is recursively constructed all the way up to the root of the tree
by joining the states corresponding to the children vertices, while simultaneously implementing a layer corresponding to their parent. The states corresponding to the trees in the decomposition
can subsequently be joined into a single state via recursive rule number $1$. Every tree decomposition results in a different final state for the key structure $\K$.

Note that we allowed only binary trees in the tree decomposition of the graph $\tilde{G}_\K$. The reason for this is that in the recursive state preparation step $2$, we join \emph{two} states in such
a way that we can implement a \emph{binary} QKD protocol for the layer $\U_1 \cup \U_2$. However, in principle we can define a more general recursive state preparation step, in which $m$
states for sub-structures are put into a uniform superposition in a Hilbert space which corresponds to a direct sum of the original Hilbert spaces. In this way, the resulting state is an equal superposition
of $m$ states living in subspaces, which are not only orthogonal, but also locally distinguishable by every user. These can be used as $m$-dimensional GHZ states in order to generate a key in layer 
$\bigcup_m \U_k$. The drawback of this recursive rule, however, is that it uses a reduction to a QKD protocol based on $m$-dimensional GHZ states, which are not known yet.
The advantage is a larger flexibility in tree decompositions of the graph $\tilde{G}_\K$---$m$-ary trees are also allowed. This can lead to a situation where    $\tilde{G}_\K$ can be decomposed
into a smaller number of trees than with binary trees only (see  Figure \ref{fig:PSTD} (c-f)).

In what follows, we give an example of the fact that the dimension-rate trade-off can scale exponentially.
Consider $n$ users $\U_n$ and  a layered key structure $\K = \{\{n,n-1\},\{n,n-1,n-2\},\dots\{n,n-1,n-2,\dots,1\}\}$.
Using only the recursive rule $1$ to construct the corresponding state results in a local dimension
$2^{n-1}$ for users $u_n$ and $u_{n-1}$, since both of them are present in each 
of the $n-1$ layers. Additionally, this results in local dimension $2^i$ for the other users
$u_i$, since each of them is present in exactly $i$ layers. On the other hand, a state for this key structure can also be obtained by applying only the trade-off rule, by adding
the layers together with an empty key structure, starting from the smallest to the largest. Such a state has a local dimension $n$ for users $u_n$ and $u_{n-1}$ and $i+1$ for every other user $u_i$.
The price to pay is the exponential decrease of the rates. While the first state achieves a rate $1$ for every layer,
the second state achieves a rate $\frac{1}{2^{n-i}}$ for a layer of size $i$. 

Note that even this implementation offers an advantage compared to the $GHZ$ implementation explored in section \ref{sec2}. 
Noting that the local dimension of the user $u_1$ is $2$, it is clear that only a \emph{qubit} $n$-partite system $GHZ$ state can be distributed to the layer of size $n$ in each time slot. 
Therefore, in order to achieve the rate equal to $1$ in the layer $\{u_1,\dots,u_n\}$, all the time slots need to be devoted to 
the distribution of the $GHZ$ state shared among all the users.
This fact results in all the other rates being equal to $0$. On the other hand, in the implementation using the full
trade-off state, the sum of the  remaining rates quickly approaches $1$ as the number of users $n$ approaches infinity.

Let us conclude this section by a short summary of the main ideas about distributing secure keys among users of a quantum network equipped with high-dimensional multi-partite entanglement sources.
We have presented three general ideas about encoding secure key structures in such states, each of which can be analyzed by a reduction to protocols using GHZ states.
The first method can be seen as a standard solution and simply uses classical mixtures of GHZ states, where the mixture is known to the users. It uses a corresponding GHZ state  for each key in the key structure. The second method utilizes the high-dimensional multipartite
structure and uses a tensor product of the GHZ states, again one for every key in the structure, and encodes them simultaneously in high local dimensions. A protocol using this idea is presented in 
section \ref{sec1}.  The third method uses
direct sum of Hilbert spaces in order to create locally distinguishable superpositions of states implementing sub-structures. As a byproduct, such superpositions can in some sense be used as a GHZ state for
a layer containing all users of the sublayers---this is a basis for the trade-off recursive rule presented in section \ref{sec3}. 

Each of these implementations has its pros and cons. The first one achieves the worst key rates, but unlike the other two it can be used with active routing of qubit entanglement sources. The second 
one achieves excellent rates, however it requires very high local dimensions, i.e.~scaling exponentially in the number of layers. The third one can be used to supplement the second method in order to reduce the local dimensions to a linear scaling in layers, albeit at the expense of decreased key rates. 

\section{Conclusions}\label{sec:conclusions}
As quantum technologies develop, network architectures involving multiple users are becoming an increasing focus of quantum communication research \cite{B:auml2017,McCutcheon2016}. For this purpose, it is vital to know the limitations and more importantly the potential of multipartite communication protocols. We contribute to this effort by providing a straightforward protocol that makes use of recent technological advances in quantum photonics \cite{Malik:2016gua,Erhard2017}. Layered quantum communication makes full use of the entanglement structure, and provides secure keys to different subsets of parties using only a single quantum state. If the production of such states becomes more reliable, this has the potential to greatly simplify network architectures as a single source will suffice for a variety of tasks. It is known that multipartite entanglement can be recovered through local distillation procedures, even if noise has rendered the distributed state almost fully separable \cite{Huber2011}. Moreover, high-dimensional entanglement is known to be far more robust to noise than low-dimensional variants \cite{Lancien2015}, indicating that even under realistic noise, our protocols, augmented by distillation, could be applied in situations where all qubit-based protocols would become impossible. Our protocols and proofs are largely based on an extension of low-dimensional variants of key-distribution through a separation into different subspaces. We have explicitly described the protocols in non-device-independent settings (i.e. trusting the measurement apparatuses, but not the source). This is mainly due to the practical limitations of fully device-independent entanglement tests, but in principle our proposed schemes could just as well work with device-independent variants of bipartite \cite{Acin:2007db} and multipartite \cite{Pappa2012,Ribeiro2017} key distribution schemes.

While the number of quantum channel uses and the noise-resistance of entanglement scale favorably in the Hilbert space dimension, the current production rates of the proposed quantum states underlying the protocols are severely limited and exponentially decreasing in the number of parties. The central challenge in multipartite quantum communication thus still remains the identification of sources that reliably create multipartite entangled states in a controllable manner and at a decent rate. We hope that explicitly showcasing potential protocols will inspire further efforts into the production of multipartite entanglement in the lab.

\section*{Acknowledgements}
We acknowledge the support of the funding from the Austrian Science Fund (FWF) through the START project Y879-N27 and the joint Czech-Austrian project MultiQUEST (I 3053-N27 and GF17-33780L).
MP also acknowledges the support of SAIA n.~o. stipend ``Akcia Rak\'usko-Slovensko''.

\bibliography{LQKD}{}
\end{document}